\documentclass[10pt]{article}
\usepackage[letterpaper]{geometry}
\usepackage{hicss}
\usepackage{times}
\usepackage[none]{hyphenat}
\usepackage{url}
\usepackage{latexsym}
\usepackage{minted}
\usepackage{indentfirst}
\usepackage{graphicx}
\usepackage[normalem]{ulem}
\graphicspath{{images/}}
\usepackage[
    style=apa,
    backend=biber
  ]{biblatex}

\usepackage[final]{microtype}
\makeatletter
\AtBeginDocument{%
  \@ifpackageloaded{biblatex}{%
    \appto{\bibsetup}{\sloppy\emergencystretch=3em}%
  }{}%
}
\makeatother
\usepackage[normalem]{ulem}     
\usepackage[hidelinks]{hyperref}
\newcommand{\email}[1]{\uline{\href{mailto:#1}{\nolinkurl{#1}}}}
\usepackage{booktabs}
\usepackage{booktabs,array}
\usepackage{pgfplots}
\pgfplotsset{compat=1.17}  
\usetikzlibrary{calc}
\usepackage{tabularx}  
\usepackage{pgfplotstable}
\usepackage{graphicx}
\pgfplotsset{compat=1.17}
\usepackage{tikz}
\usepackage{caption}
\usepackage{array}
\usepackage{makecell}
\usepackage{xcolor}
\usepackage{soul}
\usepackage{enumitem}
\usepackage{adjustbox}
\usepackage{cancel}

\title{Beyond Training: How Workers Discover Value in Enterprise AI}

\author{Riya Sahni\\
 Columbia University\\
\email{riya.sahni@cs.columbia.edu}
\And
Lydia B. Chilton\\
Columbia University\\
\email{chilton@cs.columbia.edu}}
\date{}
\begin{document}
\maketitle
\begin{abstract}

While organizations continue to invest in enterprise AI, little is known about how individual employees find valuable use cases once these tools are deployed. We present an exploratory interview study of 10 experienced U.S. professionals using M365 Copilot and interpret accounts through Rogers' Diffusion of Innovations to examine where value appears and how use cases are found and shared. Findings reveal a strong preference for informal learning methods over structured training. No participants (0/10) reported formal training as their primary way of learning; most relied on trial-and-error (8/10) and on exchanging tips with colleagues (6/10). Participants most often used M365 Copilot for note-taking/summarization, information retrieval/explanation, and writing. They also reported perceived gains in efficiency but low confidence in mastering more advanced features. The paper discusses social learning strategies and outlines implementable steps for organizations to support the discovery of high-value use cases with available enterprise AI tools.


\end{abstract}

\subsubsection*{Keywords:}

AI productivity tools; discoverability; technology adoption

\section{Introduction}

A wave of enterprise AI tools have many companies expecting AI to
magically revolutionize work. Microsoft 365 Copilot (M365 Copilot), for example, embeds generative assistance directly into Outlook, Excel, and PowerPoint, putting AI where everyday work happens. 
Leaders who buy these AI licenses are excited about the potential to increase employee productivity, but little is known about how employees actually find valuable use cases for these tools. Unlike most technologies, enterprise AI is not designed for one specific use case - users have to figure out what to prompt, how to integrate it into their workflow, and what AI is actually capable of. Finding value in these AI tools is not automatic; it needs to be discovered.


It takes time and energy to discover specific, high-value use cases for AI in one's workflow. While formal AI trainings are widely offered, they often emphasize generic, decontextualized capabilities -- like drafting emails from bullet points -- with less focus on tailored, task-specific applications. This training format might suit traditional enterprise software, but for general-purpose enterprise AI, the ``what to use this AI feature for" depends on context and value. We ask:
\begin{enumerate}
  \setlength{\itemsep}{0pt}
  \setlength{\parsep}{0pt}
  \setlength{\parskip}{0pt}
  \setlength{\topsep}{0pt}
    \item What are the common task-level use cases for M365 Copilot in everyday work, and what value or limitations do professionals report?
    \item How do professionals discover these task-level use cases in practice, and what conditions enable discovery?
\end{enumerate}

Many adoption models discuss factors that predict \textit{whether} a user adopts and learns a technology, but we want to know \textit{how}. TAM and UTAUT are variance models that look at an individual's perception of a technology and predict intention to use, so they are better suited to predict \textit{whether} a user is inclined to adopt a technology (\cite{davis1989, venkatesh2003}). 
Recent work shows that GenAI's value is often discovered through peer interactions and iterative tinkering in real workflows (\cite{feng2024, vorvoreanu2025}). Thus, we ground our analysis in Rogers' \textit{Diffusion of Innovations} (DOI), which, consistent with the early patterns in our interviews, looks at adoption beyond individual perceptions / behavior, to the social processes of communication and peer influence that shape \textit{how} employees find value in enterprise AI (\cite{rogers2003diffusion}). 

We conducted an exploratory study to elicit concrete use cases, discovery moments, and learning choices from experienced professionals. Then, we synthesized participant patterns, grounding our analysis in Rogers' framework to answer our research questions. Three themes emerged while we examined how participants discovered valuable use cases for AI at work:
\begin{itemize}
  \setlength{\itemsep}{0pt}
  \setlength{\parsep}{0pt}
  \setlength{\parskip}{0pt}
  \setlength{\topsep}{0pt}
    \item \textbf{A preference for non-traditional learning:} despite recognizing the value of formal training, participants overhelmingly favor self-directed, experiential, and social learning methods.
    \item \textbf{Discoverability as a reactive process:} users primarily discover M365 Copilot’s capabilities through ad hoc, informal channels, like by observing colleagues' demos or via spontaneous online searches.
    \item \textbf{Efficiency-confidence gap:} although most participants report high perceived efficiency gains from using M365 Copilot, a significant gap exists in their confidence in mastering its advanced features.
\end{itemize}


The paper discusses social learning strategies and translates findings into design suggestions aligned with DOI: make wins visible and shareable (observability), lower the cost of tinkering (trialability), and embed in-product, progressive guidance for advanced tasks (reduce perceived complexity).

\section{Related Work}

\subsection{Theory Approaches for Workplace Technology}

 Rogers' \textit{Diffusion of Innovations} (DOI) (\cite{rogers2003diffusion}) is a good model for studying \textit{how} people adopt and learn new technologies in collaborative environments, because it treats adoption as a social process that unfolds through communication networks over time (\cite{Greenhalgh, Dearing2018Diffusion, FreiLandau2022MobileLearning}). The model specifies five attributes that shape how an innovation spreads in a social system: \textit{Relative advantage} is the extent to which the new technology is seen as better than current tools; \textit{compatibility} is its fit within existing routines, norms, and tools; \textit{complexity} is how difficult it feels to understand and use; \textit{trialability} is how easily people can experiment with it on a limited basis; and \textit{observability} is how visible and discussable the results are to others. Prior work uses DOI to trace how technologies spread and are learned within organizations over time (e.g., \cite{karahanna1999, frambach2002, gallivan2001}). Building on that literature and early interview patterns, we use DOI to examine how social pathways like peer exchange and iterative tinkering surface high-value enterprise AI use cases and carry them into regular workflows.


Other technology adoption models focus on determinants of individual and behavioral intention instead of the social processes that surface to find and share valuable use cases. The Technology Acceptance Model (TAM) explains individual adoption by modeling how perceived usefulness and perceived ease of use shape behavioral intention and, in turn, use (\cite{davis1989}). 
Similarly, UTAUT predicts intention and use via performance expectancy, effort expectancy, social influence, and facilitating conditions (\cite{liangyongetal.,angelaschorr,venkatesh2003}). These frameworks are strong for explaining \textit{whether} a person intends to use a defined system but do not specify mechanisms for \textit{how} use cases are discovered, refined, and shared in day-to-day work. Because our goal is to explain \textit{how} use cases emerge and spread inside firms we use DOI rather than TAM/UTAUT.

\subsection{Informal and Social Learning in the Workplace}
Informal learning research shows that most workplace skills are acquired outside of formal training, through peer exchange and reflection (\cite{marsick1990informal, eraut2004informal, boud1985reflection}). Social learning theories explain the mechanisms: people learn by observing and modeling others' behavior (\cite{bandura1977social}) and by participating in communities where shared practices are negotiated over time (\cite{lave1991situated}). Empirical studies link these peer interactions to long-lasting skill development and behavior change in organizations (\cite{livingstone2001adults, billett2001learning}). 

Guided by these prior studies, we look for informal and social interactions that could shape adoption at work. Recent studies on organizations and GenAI report both helpful and unhelpful effects from social interactions (like peer dialogues and demos) in AI adoption (\cite{peng2023impact, kim2024creatorai}). Accordingly, we use DOI to ask whether and how these types of workplace interactions and channels operate as pathways for adoption, and whether they shift observability, trialability, or perceived complexity.

\subsection{AI Adoption At the Individual Level}

Most AI-in-work research focuses on organizational readiness, governance, or broad deployment strategies. A few comparative case studies examine structural approaches to managing AI in public organizations (\cite{neumann2024exploring, alsebaihi2020winning}). But, individual-level dynamics -- understanding how employees learn, experiment with, and co-create using AI -- remains underexplored. Recent calls in literature urge research that considers AI adoption from the viewpoint of individual workers (\cite{pencheva2020big}). In response, our research focuses on how employees in the U.S. find value in AI tools like M365 Copilot in their daily work. This approach bridges the gap between top–down organizational strategies and grassroots-level experiences, offering practical guidance for designing user-centered training and support systems.
We choose M365 Copilot -- a chat-based AI productivity tool embedded across the Microsoft Office suite -- to explore AI adoption at the individual level. Its integration makes it one of the most widely deployed enterprise AI tools in the United States (\cite{forrester2023copilot, althoff2025value}). As a result, it provides a timely and high-impact context for studying how AI is encountered, learned, and adopted in real-world professional settings.


\section{Methodology}
\subsection{Study Procedure}
\begin{figure*}[ht]
    \centering
    \includegraphics[width=0.49\textwidth]{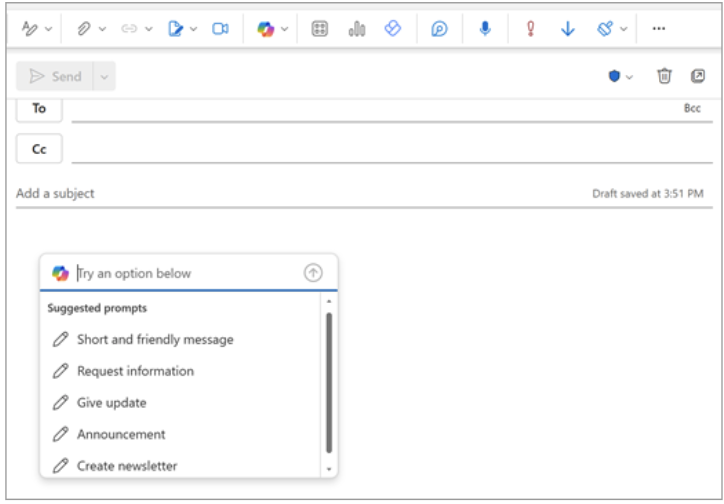}
    \hfill
    \includegraphics[width=0.49\textwidth]{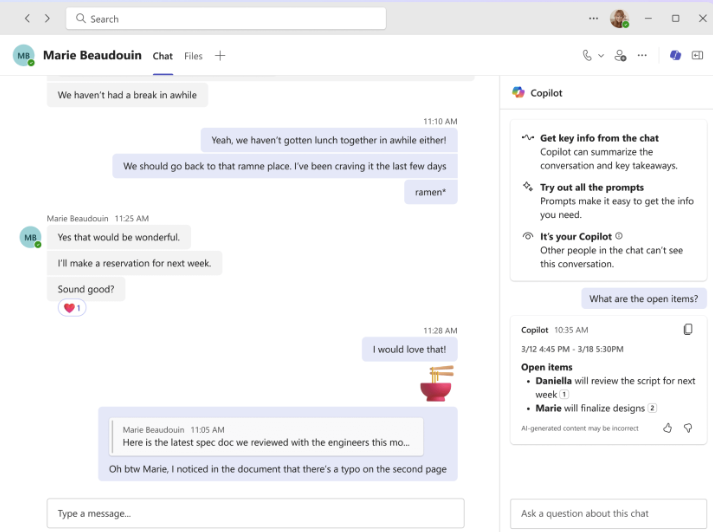}
    \caption{Example interfaces of M365 Copilot embedded into Microsoft Outlook (left) and Teams (right), as seen by study participants (from https://support.microsoft.com/).}
    \label{fig:copilot_interfaces}
\end{figure*}

Between September and December 2024, we interviewed 10 industry workers to explore their experiences with learning and using M365 Copilot tools (Figure~\ref{fig:copilot_interfaces}) at work.
The interview guideline focused on three primary themes:

\begin{itemize} 
  \setlength{\itemsep}{0pt}
  \setlength{\parsep}{0pt}
  \setlength{\parskip}{0pt}
  \setlength{\topsep}{0pt}
\item \textbf{Usage and Experience:} Participants’ daily interactions with M365 Copilot tools, including their learning strategies, feature discoverability, and both positive and negative experiences. \item \textbf{Perceptions:} Participants’ perceptions of M365 Copilot, its potential capabilities, and the extent to which these capabilities are realized in practice. \item \textbf{Learning Preferences:} Preferred learning styles and methods employed by participants to acquire knowledge about new M365 Copilot features and functionalities. \end{itemize}

Each interview lasted between 30 and 45 minutes and was conducted in English via Google Meet. All sessions were recorded and transcribed verbatim. Participants were recruited through professional networks via purposeful and snowball sampling to capture a diverse range of perspectives and behaviors. 



\subsubsection{Participant Criteria and Selection}

Participants were required to be active M365 Copilot users, engaging with the tool at least a few times per week for a minimum of two months. This ensured that participants’ usage behaviors reflected established habits rather than temporary excitement. Individuals who exclusively used other Copilot tools (like GitHub Copilot or Power BI Copilot) were excluded from the study. Focusing on M365 Copilot users was a deliberate choice, as this subscription group represents the largest population among the Copilot tools and, consequently, offers a particularly rich dataset for understanding adoption and learning behaviors.

\begin{table*}[t]
    \centering
    \caption{Participant Demographics (\textit{P prefix in ID denotes "participant"})}
    \label{tab:resized}
    \small 
    \renewcommand{\arraystretch}{1.1}  
    \setlength{\tabcolsep}{6pt}       
    \resizebox{\textwidth}{!}{ 
    \begin{tabular}{l l l c c l} 
        \toprule
        ID  & Industry                & Role                         & Experience (years) & Copilot Experience & Frequency of Usage \\
        \midrule
        P1  & Media \& Entertainment  & Senior Program Manager      & 27  & 3 months  & Daily \\
        P2  & Energy \& Chemicals     & Digital AI Lead             & 12.5  & 3 months  & Daily \\
        P3  & Energy \& Chemicals     & IT Digital Strategy Leader  & 27  & 1 year    & Daily \\
        P4  & Energy \& Chemicals     & Well Engineer               & 20  & 8 months  & Daily \\
        P5  & Transportation          & Senior Manager Analytics    & 20  & 3.5 months & Weekly \\
        P6  & Consulting              & Owner                       & 22  & 4 months  & Daily \\
        P7  & Energy \& Chemicals     & Program Manager             & 18  & 2 years   & Daily \\
        P8  & Energy \& Chemicals     & Program Manager             & 38  & 1 year    & Daily \\
        P9  & Consulting              & Senior Manager              & 12  & 1.5 years & Weekly \\
        P10 & Energy \& Chemicals     & Country Head                & 26  & 2 years   & Monthly \\
        \bottomrule
        \vspace{1mm}
    \end{tabular}
    }
\end{table*}

\subsubsection{Participant Demographics}

The sample comprises 4 women and 6 men (Table 1). All participants reported using Teams Copilot at work, while 3 also used Bing Copilot, 6 used Copilot for Outlook, 2 used Copilot for Word, and one participant also used Power BI Copilot. With the exception of two participants (P2 and P3) who have received formal training and possess extensive AI/ML experience through academic or previous professional roles, the remaining 8 participants primarily encountered AI through platforms such as Copilot, Claude, and ChatGPT. Notably, P6 leads a technical writing consulting firm that leverages various AI tools—including Quillbot and Bark—beyond the Microsoft Copilot Suite. Furthermore, P5 mentioned that she and some of her peers use Amazon Q in conjunction with M365 Copilot at work.

\subsection{Qualitative Coding \& Analysis}
The transcriptions were manually cleaned and reviewed by two researchers to eliminate transcription errors and filler words while still preserving the participants’ sentiments. Given that our objective was to examine how participant responses align with established learning frameworks rather than to uncover entirely new ones, we adopted a primarily deductive approach for the qualitative analysis (\cite{brennen2021qualitative}). Table 2 represents a codebook that was developed based on predetermined learning framework categories, including:
\begin{itemize}
  \setlength{\itemsep}{0pt}
  \setlength{\parsep}{0pt}
  \setlength{\parskip}{0pt}
  \setlength{\topsep}{0pt}
    \item \textbf{Positive social learning:} instances where participants described collaborative learning experiences, knowledge sharing, and engaging in AI-related discussions with colleagues. 
    \item \textbf{Negative social learning:} Instances characterized by resistance to knowledge sharing, limited access to peer discussions, or the dissemination of inaccurate information among colleagues.
    \item  \textbf{Experiential learning:} Self-directed, hands-on experimentation with M365 Copilot to acquire new skills and insights.
    \item \textbf{Traditional learning:} Engagement with formal company-provided training resources such as internal documentation or onboarding videos.
\end{itemize}

We also tracked experience dimensions to ground later analysis. \textit{Discoverability} draws from interface design principles around signifiers and affordances (\cite{norman2013}) and captures a users’ ability to independently find features without formal instruction. \textit{Feature opacity} reflects explainability concerns in AI system behavior (\cite{kaur2020}) and tracks the lack of clarity around what a feature does or how it behaves. \textit{Efficiency} reflects users’ perceptions of time or effort saved through intelligent systems (\cite{amershi2019}). \textit{Confidence} draws from work on calibrated trust and user reliance in automation (\cite{lee2004}) to track the user’s self-assessed comfort, trust, and willingness to delegate tasks to the tool.

Using the codebook, we reviewed and sorted participant quotes from each transcription into the relevant learning themes. To ensure coding accuracy and consistency, each transcript was reviewed twice by two researchers. Ambiguous responses were either bolstered with additional context or flagged for further discussion. If sufficient context could not be found to confidently assign a quote to a specific category, that quote was excluded from the final analysis. We also maintained a detailed audit trail of coding decisions. 


\begin{table*}[t]
  \centering
  \scriptsize
  \renewcommand{\arraystretch}{0.9}
  \caption{Qualitative Analysis Codebook}
  \label{tab:qualitative_codebook_slim}
  \begin{tabularx}{\textwidth}{@{} l X X @{}}
    \toprule
    \bfseries Code Name & \bfseries Definition & \bfseries Inclusion Criteria \\
    \midrule
    Positive Social Learning &
      Gaining knowledge by observing, mimicking, or speaking with colleagues. &
      \begin{itemize}[nosep,leftmargin=1em,label=\textbullet]
        \item Mentions learning from colleagues
        \item Appreciation for shared workplace knowledge
      \end{itemize} \\
    Negative Social Learning &
      Social learning experiences that proved ineffective. &
      \begin{itemize}[nosep,leftmargin=1em,label=\textbullet]
        \item Complaints about limited knowledge sharing
        \item Resistance to peer exchanges
        \item Colleagues providing inaccurate information
      \end{itemize} \\
    Experiential Learning &
      Hands-on experimentation to learn Copilot features. &
      \begin{itemize}[nosep,leftmargin=1em,label=\textbullet]
        \item Trial-and-error exploration
        \item Testing features without formal guidance
      \end{itemize} \\
    Traditional Learning &
      Use of formal resources like documentation or training videos. &
      \begin{itemize}[nosep,leftmargin=1em,label=\textbullet]
        \item Watching company-published tutorials
        \item Reading official documentation
      \end{itemize} \\
    \bottomrule
  \end{tabularx}
\end{table*}

\section{Findings}
\subsection{General Use Cases, Sentiments, and Usability Barriers in M365 Copilot Adoption}
\begin{figure*}[t] 
\centering
\includegraphics[width=0.8\linewidth]{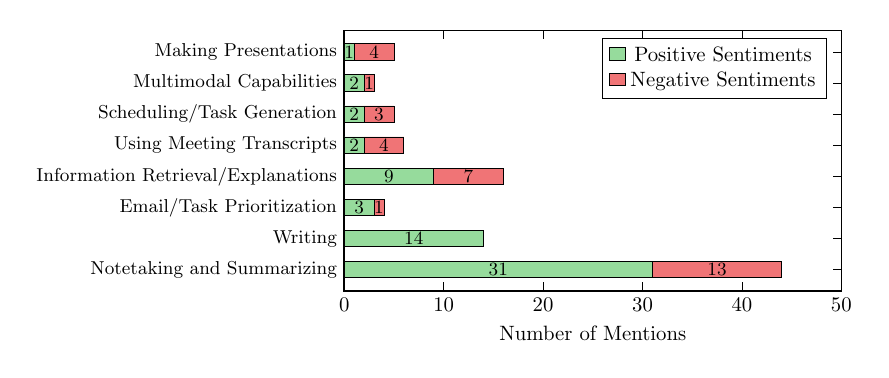}
\caption{Distribution of positive and negative sentiments in Copilot use cases}
\label{fig:sentiment_chart}
\end{figure*}

After interviewing participants, we identified eight primary use cases for M365 Copilot and categorized each mention as positive or negative sentiment based on whether participants found Copilot helpful or frustrating for those tasks (see Figure \ref{fig:sentiment_chart}).  

\subsubsection{What Participants Enjoyed About M365 Copilot}

Among 64 positive mentions, the most positively received functionalities were writing assistance (100\% positive sentiment, 14 mentions), notetaking \& summarization (70.5\% positive, 31 positive out of 44 mentions), and information retrieval/explanations (56.3\% positive, 9 out of 16 mentions). Other use cases, like email/task prioritization (75.0\% positive, 3 out of 4 mentions), meeting transcript verification (33.3\% positive, 2 out of 6 mentions), scheduling/task generation (40.0\% positive, 2 out of 5 mentions), multimodal capabilities (66.7\% positive, 2 out of 3 mentions), and presentation assistance (16.7\% positive, 1 out of 6 mentions), were mentioned less frequently and with lower positivity rates. Writing assistance was the highest-rated use case. Participants used M365 Copilot to refine emails and reports and adjust tone. P3 noted,
\begin{quote}
    ``I sometimes have the tendency to write more than I need to, and when I learned that it can do an effective summary for a wider audience, I basically started writing emails without paying a lot of attention to what I'm writing and then just dumping it into Copilot.''
\end{quote}

P4 emphasized tone adjustment capabilities, \begin{quote}
    ``Now we can adjust the tones quite a bit just by selecting if it is a casual tone or more professional tone or if it needs to be concise.''
\end{quote}

Notetaking and summarization was the most frequently mentioned positive use case overall. P5 highlighted, \begin{quote}
    ``I would say, the most memorable experience would be using Copilot to recap a meeting that I joined in 50 minutes late on a one-hour meeting.''
\end{quote} P8 also found summarization helpful:
\begin{quote}
    ``... I'm able to track the volume of information in the meetings better because I can ask Copilot to go tell me what was said in this meeting…''
\end{quote}




Overall, participants appreciated how M365 Copilot streamlined straightforward tasks like email writing and meeting summarization.


\subsubsection{Issues That Participants Faced With M365 Copilot} 

While every participants claimed to find M365 Copilot useful for at least one task, 33 total negative mentions also emerged (Figure ~\ref{fig:sentiment_chart}). The most criticized features were presentation assistance (66.7\% negative) and meeting transcript verification (66.7\% negative). Notetaking and summarization (29.5\% negative) and information retrieval (43.8\% negative) were frequently used but also generated frustration. For presentation assistance, P7 commented:
\begin{quote}
    ``The PowerPoint outputs felt like a generic Google answer, not something tailored to our work.''
\end{quote}
Transcript summarization accuracy was also a concern. P10 noted:
\begin{quote}
    ``The Copilot transcript summary didn't quite capture the essence of what people were saying.''
\end{quote}
P1 recounted missing important discussion points:
\begin{quote}
    ``There was one meeting. And I'm glad I looked at the notes relatively early because there was a whole section of a discussion that was really important. And it wasn’t included in the Copilot notes at all.''
\end{quote}
P5 described how inaccuracies led her team to avoid using Copilot for technical discussions:
\begin{quote}
    ``We have stopped using [Teams Copilot] to capture anything for code reviews, or anything at all for technical now based on that one experience because we don’t want to get it wrong.''
\end{quote}

Some participants also experienced irrelevant search results when retrieving information. P2 shared:
\begin{quote}
    ``I asked Copilot to retrieve past meeting decisions, but it said it had no recollection. I manually found the transcript with the needed information.''
\end{quote}

Other frustrations were reported across smaller features like scheduling, automation, and data parsing. P6 described, ``I had a workflow set up to automate table sharing, but it kept glitching.'' P5 noted,
    ``If the table name was called A\_B, it incorrectly merged it into ABCD.''
Overall, participants' primary issues with M365 Copilot centered on inaccuracies and inconsistencies in its more advanced features, like presentation assistance and meeting transcript verification. 
\\ \\

\subsection{High Perceived Efficiency using M365 Copilot at Work but Low Confidence in its Mastery}

\begin{figure}[ht]
\centering
\begin{tikzpicture}
\definecolor{red}{RGB}{240,116,118}
\definecolor{green}{RGB}{150,219,156}

\begin{axis}[
    xlabel={Perceived Efficiency Rating},
    ylabel={Perceived Confidence Rating},
    xmin=0, xmax=10,
    ymin=0, ymax=10,
    xtick distance=1,
    ytick distance=2,
    grid=both,
    width=1.07\linewidth,
    height=6cm,
    title={Efficiency vs. Confidence Ratings for M365 Copilot},
    title style={font=\small},
    label style={font=\small},
    tick label style={font=\small}
]

\addplot [domain=0:10, dashed, color=gray] {x};

\addplot+[
    scatter, only marks, mark=*, color=black,
    visualization depends on={value \thisrow{label} \as \labelname},
    nodes near coords={\labelname},
    nodes near coords style={
        anchor=south west, 
        font=\tiny, 
        fill=white, 
        inner sep=0.5pt, 
        xshift=1.5pt, 
        yshift=1.5pt
    },
    mark options={fill=red, draw=red}
]
table[meta=label, row sep=\\] {
x y label \\
5 8 P5 \\
6 8 P7 \\
3 5 P10 \\
};

\addplot+[
    scatter, only marks, mark=*, color=black,
    visualization depends on={value \thisrow{label} \as \labelname},
    nodes near coords={\labelname},
    nodes near coords style={
        anchor=south west, 
        font=\tiny, 
        fill=white, 
        inner sep=0.5pt, 
        xshift=1.5pt, 
        yshift=1.5pt
    },
    mark options={fill=green, draw=green}
]
table[meta=label, row sep=\\] {
x y label \\
4 2 P1 \\
7 5 P2,P9 \\
8 5 P3,P4 \\
9 1 P6 \\
9 3 P8 \\
};

\end{axis}
\end{tikzpicture}
\caption{Perceived Efficiency and Confidence ratings across participants. Red dots indicate confidence exceeds efficiency; green dots indicate confidence is lower or equal to efficiency.}
\label{fig:scatter}
\end{figure}

While every participant perceived M365 Copilot to improve efficiency in at least one task, more than half expressed low confidence in using it effectively, suggesting a gap between perceived usefulness and perceived complexity. Seven out of ten participants rated Copilot’s impact on their efficiency as 6 or higher on a 10-point scale (Figure \ref{fig:scatter}). Only 2 rated their confidence in mastery at 7 or above. P8 rated efficiency a 9 and confidence at 3, explaining: 

 
\begin{quote}
    “Using it in Teams just makes me, I don’t want to say smarter, but it makes me more engaged and helps me understand more about what’s going on.” 
\end{quote}
Several participants reported actively trying multiple learning approaches (like workshops, documentation, peer exchanges), but still felt unsure whether they were accessing the full range of features. P5 expressed that even after attending formal trainings, she still did not know what new features were available. Part of this confidence gap stemmed from how quickly M365 Copilot's capabilities evolved. P7 described, \begin{quote}
    “Even the name changing of the software and the lineage of the applications is really difficult for people who aren’t deeply engaged. It almost becomes an obsession to stay connected with what’s happening.”
\end{quote}

Overall, participants consistently experienced M365 Copilot as both immediately helpful and difficult to fully master as its features shifted over time.

\subsection{Low Trialability and High Perceived Complexity Led Users to Abandon Copilot Early}


Many participants explored M365 Copilot through trial-and-error but often abandoned use quickly when results were unsatisfactory. These patterns highlight low trialability and high perceived complexity, which DOI identifies as barriers to sustained adoption.

P9 explained: \begin{quote}
    ``If I need to do something and I can't find an answer, I try to look up videos and do it. But if I don’t need anything, I don’t know what else it can do. You don’t know what you don’t know.''
\end{quote} 

 Participants often cited time contraints and competing priorities as the main reasons for avoiding formal training. P8 described his limited exploration window:

\begin{quote}
    ``I didn’t spend a whole lot of time [trying to get Copilot to generate a PowerPoint presentation]... it was probably 15 minutes or so.'' 
\end{quote}


Others expressed a tendency to abandon tasks if Copilot failed early in the process. P7 summarized:
\begin{quote}

    ``if you're not going in the direction that feels like it's going to be successful,... you abandon that use case and you move on... you just say `I'm not going to waste any more time with this... I'd rather do it manually'.''

\end{quote}

Across participants, underutilization appeared driven less by lack of access to training, and more by competing priorities, limited patience, and absence of clear incentives to explore additional features.



\subsection{Experimentation and Observation Drove Adoption More than Training}

Participants repeatedly described discovering M365 Copilot's capabilities by watching colleagues use it or by directly exchanging prompts. These peer interactions increased the \textit{observability} of AI capabilities. They also compensated for low formal trialability and high perceived complexity.

\subsubsection{Participants Ignored Formal Trainings.}


Although 9 out of 10 participants acknowledged formal training existed, seven intentionally skipped official onboarding materials. Sixty percent preferred self-guided, hands-on learning through trial and error and social learning over formal training. Five participants cited time constraints as a major barrier for going through formal trainings as a primary learning method. 

Participants often leaned on peer interactions instead. P4 described casually teaching colleagues:

\begin{quote}
     ``... and before lunch was even over, I sent the job description out, and they’re like, ‘How did you do that?’ I said ‘it’s Copilot.’...and then I showed them. They’re like ‘wow’...''. 
\end{quote}

P5 recalled learning through peer screen-sharing sessions:
\begin{quote}
    ''I realized we could share screenshots back and forth and say, `that’s how you do that', that's `how you do a ticket integration...' we learned from each other by sharing our screen during some of the office hours...”
\end{quote}

Overall, most participants skipped formal training and instead learned by talking to colleagues and trying things out as needed.

\subsubsection{Discoverability and AI Adoption}



Sixty-percent of participants primarily learned through peer-based social learning, while 30\% relied on self-exploration. Only 1 participant had minimal engagement with M365 Copilot, instead relying on assistants. Peer demonstrations frequently served as the entry point for adoption. P8 described how observing a colleague's prompt led to personal experimentation:

\begin{quote}
    ``One of the guys ... thought he was trying to use Copilot but instead he wrote out his prompt in the chat and I saw it... so I pinged him on the side ... and he showed me some things and I’m like, `Son of a gun, all right…'''
\end{quote}

P1 described being influenced by seeing coworkers use M365 Copilot:
\begin{quote}
    ``I had peers, ... who were part of the pilot program. They would be on conference calls, running Copilot, and I was like, ‘I want that.’''
\end{quote}

P5 emphasized the motivating role of peer-driven experimentation:
\begin{quote}
     ``Our internal group learned from each other... It was peer pressure. Others were testing [Teams Copilot], so we were too. We were challenged to try new features every day. When we were on our own, like with Power BI Copilot, we weren’t as motivated.''
\end{quote}

As a result, most participants found new M365 Copilot features by watching others experimenting on their own, rather than through formal learning resources.

\section{Discussion}

\subsection{Observed Preference towards Social Learning over Formal Training for M365 Copilot}

Participants skipped M365 Copilot’s official training videos and documentation, instead learning by watching peers, swapping prompts in Teams channels, and tinkering on their own. The trialability and observability attributes in DOI help explain why. A colleague's on-screen demo provides an instant, low-risk trial: employees can copy the prompt, tweak it, and see the result inside of their own workflow in minutes. Because the \textit{gain} is public (for example, better wording and more concise emails), the benefits of this demo are visible to others, which encourages rapid imitation.

This mode of learning also mitigates \textit{perceived complexity}, because  it turns an abstract feature into a copyable, working example in a familiar tool. While formal tutorials might become outdated after new feature releases and product updates, peer exchanges update themselves organically. As soon as one user discovers a new shortcut, it spreads in chat channels. Organizations can therefore support diffusion not by producing longer training decks, but by institutionalizing these micro-communities -- perhaps by rotating a ``prompt-of-the-week" slot in stand-ups or curating a living FAQ that anyone can edit after a successful experiment.

\subsection{Feature Opacity and the Efficiency-Confidence Gap}

Every participant highlighted at least one of M365 Copilot's surface-level advantages like drafting emails in seconds, summarizing long meetings, and retrieving buried documents. These quick wins help explain why 7 of the 10 interviewees rated the tool's impact on their efficiency at 6 or higher (Figure 2). This is a classic case of high \textit{relative advantage}, as mentioned in DOI. Yet, the same people scored themselves low on mastery, and conversation after conversation would return to the refrain ``\textit{... I'm sure Copilot can do more, but I just don't know where to start.}" This hesitation stems from what our data reveal as \textit{feature opacity}.

Advanced capabilities are not just hard to find, but they can be non-intuitive to trigger. P8 and P1 incidentally discovered some M365 Copilot features by watching others use it. Feature opacity inflates perceived complexity and undercuts two more DOI attributes: trialability (because experiments with Copilot can fail fast) and observability (because successful deep-feature uses rarely become visible to peers). This can result in the efficiency-confidence gap illustrated in Figure \ref{fig:scatter}. M365 Copilot looks valuable, but it feels enigmatic.

Our self-identified ``tinkerers" (like P2 and P7) behave like \textit{innovators} in Moore's \textbf{Crossing the Chasm}: they push past the initial perceived complexity, post screenshots of successful prompts, and make their wins observable to colleagues. When their experiments circulate, others follow. When early adopters stop at surface-level tasks, diffusion stalls and the majority stay in cautious ``basic use" mode. Early-adopter activity raises observability for the \textit{early majority}, but only if organizations recognize and support it.

\subsection{Design Implications} 
Design interventions for enterprise AI tools like M365 Copilot should not just focus on expanding features, but reducing perceived complexity and encouraging more peer-observability and trialability as AI's capabilities evolve.
These implications are grounded in emergent patterns across our interviews, and should be interpreted as exploratory suggestions rather than definitive design directives.
\subsubsection{Incremental Learning Milestones}

 A few participants admitted to abandoning some features after short, unsuccessful trails, while others reported relying mostly on basic use cases like drafting emails and summarizing meetings. This hesitancy reflects the \textit{complexity} attribute from DOI. When features feel opaque or overwhelming, users hesitate more to try them. Rather than presenting long, upfront trainings, design interventions can instead layer learning into users' natural workflows. Having in-app ``show me one more thing" buttons or progressive walkthroughs that reveal hidden options after a successful basic task can gradually build mastery in a user without overwhelming them. These incremental learning milestones lower the cognitive load, reducing perceived complexity and feature opacity.   

\subsubsection{Stimulating More Peer-Observability and Trialability}

Instead of formal training, many participants learned through Teams chats and casual screen shares. This reflects the observability and trialability attributes from DOI: visible successful prompts encouraged others to imitate and also lowered the risks of a failed experiment with M365 Copilot. Design features should allow users to easily share and test successful prompts. 
Organizations can also spotlight early adopters (e.g. 5-minute ``Copilot hacks" in team meetings) so that trialability and peer validation can both be pushed. Over time, these types of cues can move users beyond initial advantage-seeking to more exploratory use without additional formal training.

\section{Limitations and Future Work}


This is primarily an exploratory study with a sample size of only 10 participants. Thus, individual perspectives may not fully reflect broader organizational or industry-wide trends. While we included participants from diverse backgrounds and roles, future studies should incorporate larger-scale surveys or longitudinal studies to validate and expand upon these findings. Our interview recruitment methods may also introduce bias by overrepresenting users who are already engaged with AI tools and underrepresenting those who may have disengaged early. Participant demographics skewed senior and averaged over 22 years of industry experience. Future work should examine how early-in-career and digitally native users engage with AI tools like M365 Copilot, as their expectations and experimentation habits may differ. Future work should also explore how other consumer-grade GenAI tools like ChatGPT or Claude might shape user expectations, learning behaviors, and trust in enterprise-specific systems like M365 Copilot.

\section{Conclusion}

Our work identifies three key factors shaping M365 Copilot adoption: a strong preference for experiential and social learning, the role of informal discovery in learning new features, and a persistent gap between perceived efficiency gains and confidence in mastery. These insights imply that organizations should look beyond traditional training methods. Embedding real-time feedback, contextual tips, or interactive modules directly within the tool may help workers discover value in enterprise AI faster and achieve greater productivity gains.
\addtolength{\textheight}{-.2cm} 
\printbibliography










\end{document}